\documentclass[prd, aps,footinbib, preprintnumbers, showpacs,superscriptaddress,twocolumn]{revtex4}
\usepackage{latexsym}
\usepackage{amssymb}
\usepackage{amsfonts}
\usepackage{amsmath}
\usepackage[dvips]{graphicx}\usepackage{color}

\usepackage{ulem}

\renewcommand{\emph}[1]{{\it #1}}
\newcommand{\Ref}[1]{(\ref{#1})}
\newcommand{\be}{\begin{equation}}  
\newcommand{\ee}{\end{equation}}
\newcommand{\bea}{\begin{eqnarray}}           
\newcommand{\eea}{\end{eqnarray}} 
\newcommand{\beqn}{\begin{eqnarray*}}
\newcommand{\eeqn}{\end{eqnarray*}}
\newcommand{\ba}{\begin{align}}
\newcommand{\ea}{\end{align}}

\def\p4{{\psi_4}} 

\begin{document}


\title{An improved analytical description of inspiralling and coalescing black-hole binaries}

  \author{Thibault \surname{Damour}}
  \affiliation{Institut des Hautes Etudes Scientifiques, 91440 Bures-sur-Yvette, France}
  \affiliation{ICRANet, 65122 Pescara, Italy}

  \author{Alessandro \surname{Nagar}}
  \affiliation{Institut des Hautes Etudes Scientifiques, 91440 Bures-sur-Yvette, France}
  \affiliation{ICRANet, 65122 Pescara, Italy}
  \affiliation{INFN, Sezione di Torino, Via Pietro Giuria 1, Torino, Italy}

  \date{\today}
  
\begin{abstract}
We present an analytical formalism, within the Effective-One-Body
framework, which predicts gravitational-wave signals from inspiralling and
coalescing black-hole binaries that agree, within numerical errors,
with the results of the currently most accurate numerical relativity simulations
for several different mass ratios. In the equal-mass case, the 
gravitational wave energy flux
predicted by our formalism agrees, within numerical errors, with the most
accurate numerical-relativity energy flux.
We think that our formalism opens a realistic
possibility of constructing a sufficiently accurate, large bank of gravitational
wave templates, as needed both for detection and data analysis of (non spinning)
coalescing binary black holes.

\end{abstract}

  \pacs{
    04.25.Nx,   
    04.30.-w,   
    04.30.Db    
  }
  
  \maketitle

The opening of gravitational wave (GW) astronomy depends on our theoretical ability at computing, within Einstein's theory
of General Relativity, a sufficiently accurate approximation
to the GW signal emitted by the premier target of ground-based 
GW detectors: inspiralling and coalescing binary black holes (BBH's).
Indeed, to detect these GW signals, and extract physical information
from them, one must correlate the noisy output of the detectors to 
a very large bank of ``GW templates'', each template giving an 
accurate representation of the gravitational waveform emitted 
by a BBH with certain physical parameters (notably masses and spins).
Recent breakthroughs in 
Numerical Relativity 
(NR)~\cite{Pretorius:2005gq,Campanelli:2005dd,Baker:2005vv,Gonzalez:2006md,Koppitz:2007ev,Scheel:2008rj} 
have given us access to an accurate knowledge of the 
waveform emitted during the late inspiral and merger 
of a sparse sample of BBH systems
(see~\cite{Pretorius:2007nq} for a review). 
However, BBH simulations are time consuming. 
This precludes the sole use of NR simulations
for building the needed bank of GW templates, densely filling
the multidimensional space of BBH physical parameters (masses
and spins). It is urgent to have in hands an {\it analytical
method} able to give a sufficiently accurate representation of
the motion of, and the gravitational radiation from, 
coalescing binary black holes with arbitrary masses and 
spins. We shall describe here a formalism which hopefully 
solves this problem, in the case of circularized, 
non-spinning binary black holes with arbitrary masses $m_1$, $ m_2$.

The analytical formalism presented here is a significantly 
improved version of the general Effective-One-Body (EOB) 
method~\cite{Buonanno:1998gg,Buonanno:2000ef,Damour:2000we,Damour:2001tu}. 
The predictions of previous (less accurate) implementations 
of the EOB method have already been compared, with success, 
to various types of results from 
numerical simulations~\cite{Damour:2002qh,Buonanno:2006ui,Damour:2007cb,Pan:2007nw,
Damour:2007xr,Buonanno:2007pf,Damour:2007yf,Damour:2007vq,Damour:2008te,Boyle:2008ge}. 
Before explaining the improvements that characterize
our  formalism, let us recall that the {\it four } essential 
elements of the EOB approach
are:
(i) a {\it Hamiltonian} $H_{\rm real}$ describing the conservative 
part of the relative dynamics of the two
black holes; (ii) a {\it radiation-reaction force} ${\mathcal F}_{\varphi}$
describing the loss of
(mechanical) angular momentum, and energy, of the binary system; 
(iii) the definition of the various multipolar components of the {\it``inspiral-plus-plunge''
(metric) waveform} $h_{\ell m}^{\rm insplunge}$; and
(iv) the attachment of a subsequent {\it ``ringdown waveform''} $h_{\ell m}^{\rm ringdown}$ around a
certain (EOB-determined)``merger time'' $t_m$. The latter
fourth facet of the EOB formalism, namely the assumption
of a sharp transition, around the BBH merger, between
the ``plunge'' and a ringdown behavior, was inspired 
by the classic ``plunging test-mass'' result of~\cite{Davis:1972ud}.
This assumption has been well confirmed by the 
results of NR simulations~\cite{Pretorius:2007nq},
and we shall not try here to improve on it.
We shall follow here the usual EOB 
procedure~\cite{Buonanno:2000ef} of matching 
the ``insplunge'' and ``ringdown'' waveforms 
around the EOB merger, and matching, time $t_m$, 
defined as the location of the maximum EOB orbital 
frequency. [We use the specific 
procedure of~\cite{Damour:2007xr,Damour:2007vq}, 
with 5 quasi-normal-modes, and a ``comb'' of 
spacing $\delta=2.3 M_{\rm f}$, where $M_{\rm f}$ is the mass 
of the final black hole.]

\begin{figure*}[t]
\begin{center}
\includegraphics[width=120 mm, height=45mm]{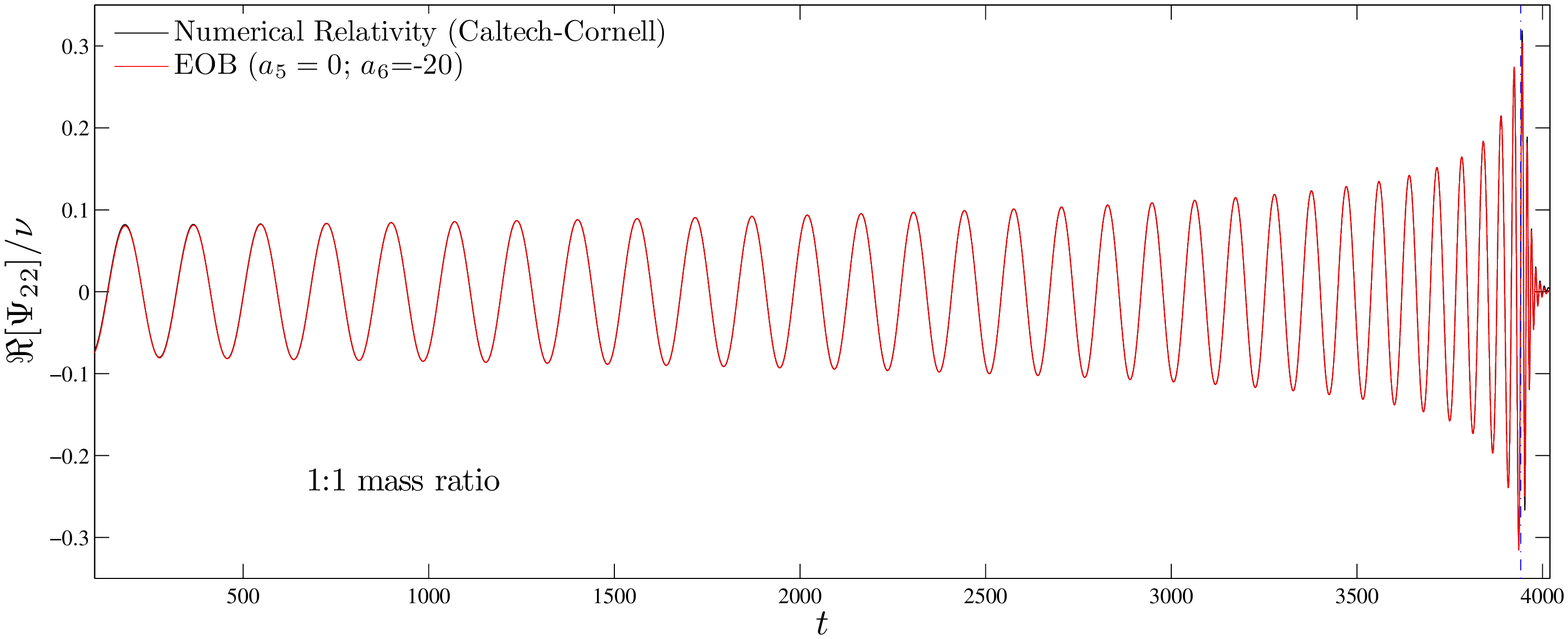}
\hspace{-12 mm}
\includegraphics[width=60 mm, height=45mm]{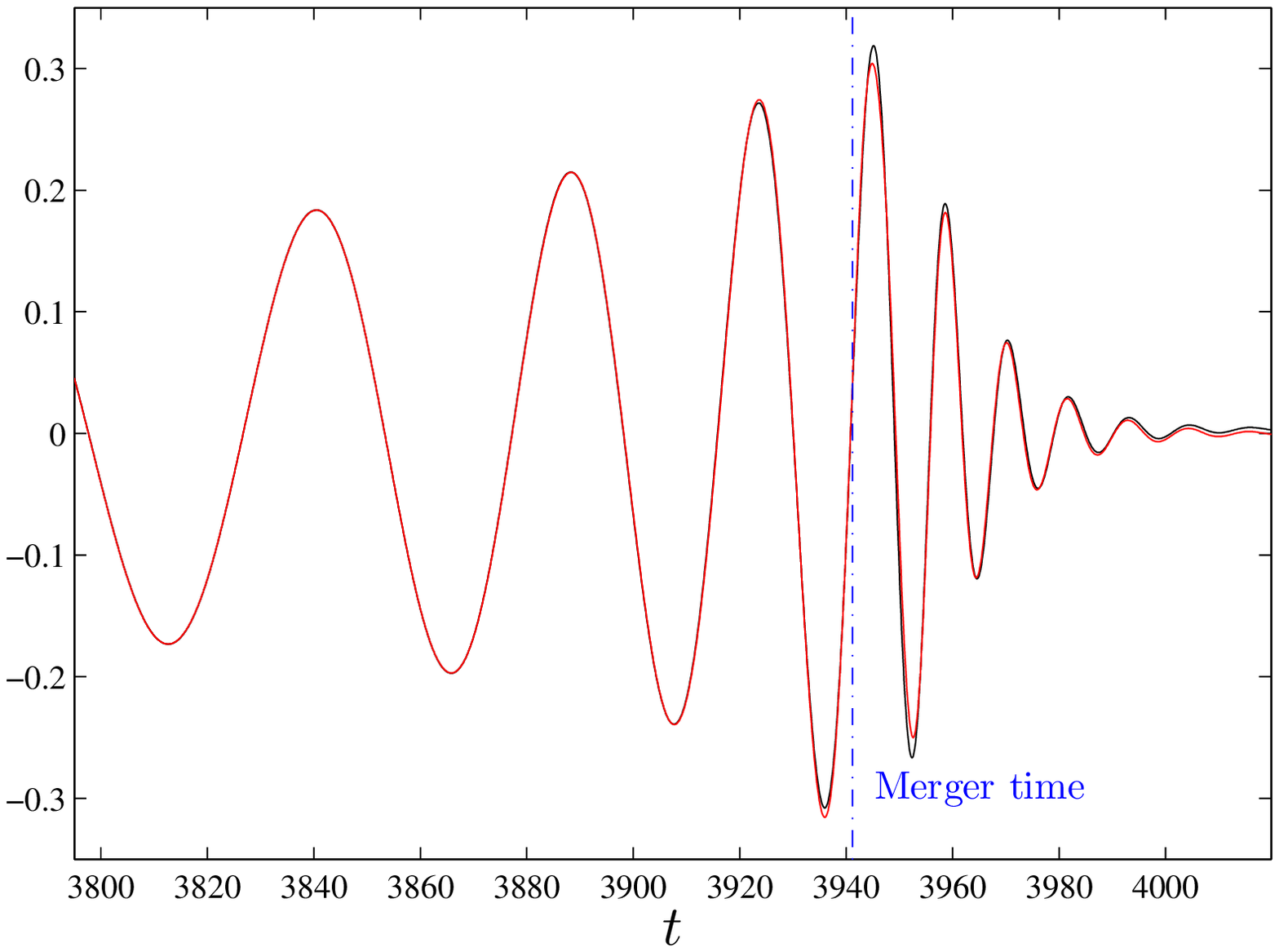}
\caption{ \label{fig:fig1} Equal-mass case: agreement between 
NR (black online) and EOB-based (red online) $\ell=m=2$ metric waveforms.}
\end{center}
\hspace{-5mm}
\end{figure*}
On the other hand, we propose here to improve the three 
other basic elements of the EOB formalism in the following way:

(i) The central object entering the relative Hamiltonian,
$H_{\rm real} = M [ 1 + 2\nu( \hat{H}_{\rm eff} - 1)]^{1/2}$,
where $M \equiv m_1 + m_2$, $\mu  \equiv m_1 m_2/M$, 
$\nu  \equiv \mu/M$, $z_3= 2 \nu (4-3\nu)$, and
\begin{equation}
\label{eq:Heff}
\hat{H}_{\rm eff}\equiv \sqrt{ p_{r_*}^2 + A(1/r)\left(1+{p_{\varphi}^2}/{r^2}\
+z_3{p_{r_*}^4}/{r^2} \right)} 
\end{equation}
is the  EOB radial potential $A(u)$ (here
$u=1/r$; the EOB radial coordinate $r$ is rescaled by
$M$, with $G=c=1$; and $p_{r_*}$
is canonically conjugated to the EOB ``tortoise-like''
radial coordinate $r_*$, defined in~\cite{Damour:2007cb}).
Current analytical calculations, within the post-Newtonian (PN)
formalism, of the dynamics of BBH's have computed the first
four terms (3PN approximation) in the (Taylor) expansion
of the radial potential $A(u)$ in powers of 
$u$~\cite{Damour:2000we,Damour:2001bu}, namely
$A^{\rm 3PN}(u) \equiv 1- 2u + 2\nu u^3 +\left(94/3- 41\pi^2/32\right) \nu u^4 $.
Here we propose to consider the two-parameter class
of (extensions and) resummations of  $A^{\rm 3PN}(u)$ 
defined by
\be
\label{A}
A(u;a_5,a_6,\nu) \equiv P^1_5[A^{\rm 3PN}(u) + \nu a_5 u^5 + \nu a_6 u^6],
\ee
where $P^1_5$ denotes a  $(1,5)$ {\it Pad\'e approximant}.
This is a generalization of the one-parameter ($a_5$)
class of $P^1_4$ Pad\'e-resummed $A$-potentials considered
in previous EOB works~\cite{Damour:2001tu,Buonanno:2007pf,Damour:2007yf}.

(ii) The second (and most novel) improvement that we
introduce here concerns the radiation reaction force
${\mathcal F}_{\varphi}$. We make use of the
very recent results of ~\cite{Damour:2008gu} concerning
an ``improved resummation'' of post-Newtonian 
multipolar waveforms. Specifically, we define
${\mathcal F}_{\varphi}$ in the following way
($\Omega$ denoting the EOB orbital frequency):
\begin{equation}
\label{RR}
{\mathcal F}_{\varphi} \equiv - \frac{1}{8 \pi \, \Omega} \sum_{\ell
= 2}^{\ell_{\rm max}} \sum_{m=1}^{\ell} (m \, \Omega)^2 \, \vert R \, h_{\ell m}^{(\epsilon)} \vert^2 \, .
\end{equation}
Here, we shall take $\ell_{\rm max} = 8$ and define the individual
multipolar waveforms $h_{\ell m}^{(\epsilon)}$
(where $\epsilon = 0,1$ labels the ``even'' or ``odd'' parity) 
in the following way: (a) the leading quadrupolar contribution to 
${\mathcal F}_{\varphi}$, i.e. the term $\epsilon = 0$ and $\ell =m= 2$ 
in \Ref{RR}, is computed by using the quadrupolar
waveform defined in Eq.~\Ref{h22} below; while (b) the subdominant 
terms (i.e. when, either $\epsilon = 1$, or $\epsilon = 0$ 
and $\ell \geq 3$), $h_{\ell m}^{(\epsilon)}$ are defined 
by Eq.~(1) of~\cite{Damour:2008gu}, together with the other
definitions  given there (see~\cite{DN} for details).

(iii) The third improvement introduced here concerns the
(dominant) even-parity, quadrupolar ($\epsilon = 0$ and $\ell =m= 2$) 
``insplunge'' waveform. We take it in the form
\begin{equation}
\label{h22}
h_{22} = \frac{M\nu}{R} \, n_{22} \, x \, Y^{2,-2} 
\left( \frac{\pi}{2} , \Phi \right) \hat H_{\rm eff} \, T_{22} \, e^{i \delta_{22}}  f_{22}(x) \, f_{22}^{\rm NQC} \, ,
\end{equation}
where, for notational simplicity, we have 
suppressed the parity label $\epsilon = 0$.
Such a {\it multiplicatively decomposed} form of $h_{22}$
was introduced in \cite{Damour:2007xr,Damour:2007yf}. 
See these references, and~\cite{Damour:2008te,Damour:2008gu,DN} 
for the definition of the factors in \Ref{h22}; 
$f_{22} (x)$ is a modulus correction, here defined 
(as in \cite{Damour:2007yf}, where $f_{22}$ was computed, 
using~\cite{Blanchet:2004ek,Tanaka:1997dj},
at the $3^{+ 2}$ PN accuracy; Eq.~(10) there) as being 
the following {\it Pad\'e-resummed} function
$
f_{22}^{\rm Pf} (x;\nu) = P_2^3 [f_{22}^{\rm Taylor} (x;\nu)].
$
The new ingredient of $h_{22}$ introduced here
is the definition of the last factor in Eq. \Ref{h22},
namely an additional ``Next-to-Quasi-Circular'' (NQC) correction 
factor of the form\footnote{Note that one could also 
similarly improve the subleading higher-multipolar-order 
contributions to ${\mathcal F}_{\varphi}$. We leave the 
exploration of such  refinements to future work.}
\begin{equation}
\label{fNQC}
f_{22}^{\rm NQC} (a_1 , a_2) = 1 + a_1 \, p_{r_*}^2/(r\Omega)^2 + a_2 \, \ddot r / r \, \Omega^2 \, .
\end{equation}
A crucial facet of the new analytical formalism presented here
consists in trying to be as predictive as possible by
reducing to an absolute minimum the
number of ``flexibility parameters'' entering our theoretical 
framework. We shall achieve this aim by ``analytically'' 
determining  the two parameters $a_1, a_2$
entering (via the NQC factor Eq.~\Ref{fNQC})
the (asymptotic) quadrupolar EOB waveform 
$\hat R h^{\rm EOB}_{22}$ (where $\hat R =R/M$)
by imposing: (a) that the modulus $|\hat R h^{\rm EOB}_{22}|$
reaches, at the EOB-determined ``merger time'' $t_m$, a
{\it local maximum}, and (b) that the value of this
maximum EOB modulus is equal to a certain 
(dimensionless) function of $\nu$, $\varphi(\nu)$.
We  calibrated $\varphi(\nu)$ (independently 
of the EOB formalism) by extracting from the 
best current Numerical Relativity simulations 
the maximum value of the modulus of the 
Numerical Relativity quadrupolar
{\it metric} waveform $|\hat R h^{\rm NR}_{22}|$.
Using the data reported in \cite{Scheel:2008rj} and
\cite{Damour:2008te}, and considering the 
``Zerilli-normalized'' asymptotic metric 
waveform $\Psi_{22} = \hat R h_{22}/\sqrt{24}$, we found
$\varphi(\nu) \simeq 0.3215 \nu ( 1 - 0.131 (1-4\nu))$.
Our requirements (a) and (b) impose,
for any given $A(u)$ potential, {\it two constraints} on the
{\it two parameters} $a_1, a_2$. We can solve these two
constraints (by an iteration procedure) and thereby uniquely
determine the values of $a_1, a_2$ corresponding to any
given $A(u)$ potential. In particular, in the case
considered here where $A(u;a_5,a_6,\nu)$ is defined by
Eq. \Ref{A}, this uniquely determines $a_1, a_2$
in function of $a_5,a_6$ and $\nu$.

Finally, our analytical formalism contains {\it only two analytically
undertermined parameters}, namely $a_5$ and $a_6$, which parametrize 
some flexibility in the Pad\'e-resummation of the basic 
radial potential $A(u)$, connected to the yet uncalculated 
higher PN contribution~\footnote{Indeed, $a_5$ and $a_6$ 
formally parametrize the 4PN and 5PN contributions to the 
Taylor expansion of $A(u)$.}.
\begin{figure}[t]
\begin{center}
\includegraphics[width=70 mm, height=55mm]{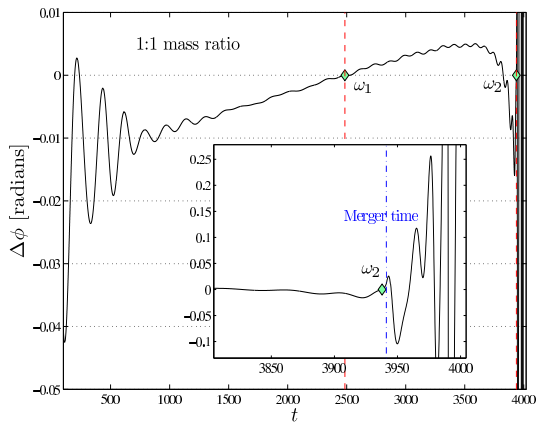}
\caption{ \label{fig:fig2} Phase difference between 
the analytical and numerical (metric) waveforms of Fig.~\ref{fig:fig1}.}
\vspace{-5mm}
  \end{center}
\end{figure}
We have first compared the $(a_5,a_6)$-dependent predictions
made by our formalism to the high-accuracy waveform from an
equal-mass BBH ($\nu=1/4$) computed by the Caltech-Cornell 
group~\cite{Scheel:2008rj} (and now made available on the web).
We found that there is a strong degeneracy between $a_5$ and $a_6$
in the sense that there is an excellent EOB-NR agreement 
for an extended region in the
$(a_5,a_6)$-plane. More precisely, the phase difference between
the EOB (metric) waveform and the Caltech-Cornell one, considered
between GW frequencies $M\omega_{\rm L}=0.047$ and 
$M\omega_{\rm R}=0.31$ (i.e., the last 16 GW cycles before merger),
stays smaller than 0.02 radians within a long and thin 
banana-like region in the $(a_5,a_6)$-plane. This ``good region'' 
approximately extends between the points 
$(a_5,a_6)=(0,-20)$ and $(a_5,a_6)=(-36,+520)$. 
As an example (which actually lies on the boundary of the 
``good region''), we shall consider here the
specific values $a_5=0, a_6=-20$ (to which correspond, 
when $\nu=1/4$, $a_1 = -0.036347, a_2=1.2468$). 
We henceforth use $M$ as time unit.

\begin{figure}[t]
\begin{center}
\includegraphics[width=70 mm, height=55mm]{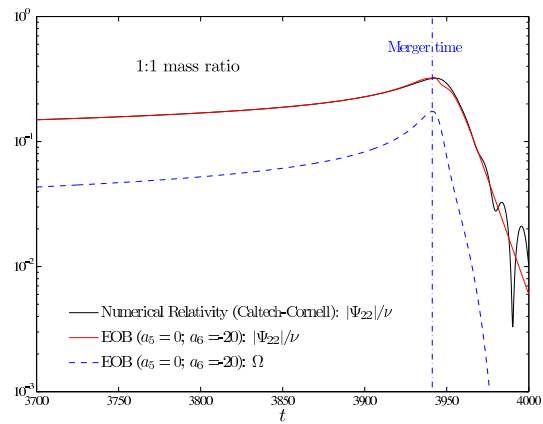}
\caption{ \label{fig:fig3}Equal mass case: metric-amplitudes comparison. 
The maximum of the orbital frequency $\Omega$ defines the EOB merger.}
\vspace{-5mm}
  \end{center}
\end{figure}

Figure~\ref{fig:fig1} compares (the real part of) our 
analytical {\it metric} quadrupolar waveform 
$\Psi^{\rm EOB}_{22}/\nu$ to the corresponding 
(Caltech-Cornell) NR {\it metric} waveform $\Psi^{\rm NR}_{22}/\nu$
(obtained by a double time-integration, \`a 
la~\cite{Damour:2008te}, from the original 
NR {\it curvature} waveform $\psi_{4}^{22}$). 
[We used the ``two-frequency pinching technique'' 
of~\cite{Damour:2007vq} with $\omega_1=0.047$ 
and $\omega_2=0.31$.]  The agreement 
between the analytical prediction and the NR 
result is striking, even around the merger 
(see the close-up on the right).
The phasing agreement is excellent over the full 
time span of the simulation
(which covers 32 cycles of inspiral and about 6
cycles of ringdown), while the modulus agreement is excellent
over the full span, apart from two cycles after merger
where one can notice a difference. A more quantitative
assessment of the phase agreement is given in Fig.~\ref{fig:fig2},
which plots the ($\omega_1$-$\omega_2$-pinched) phase
difference $\Delta \phi= \phi_{\rm metric}^{\rm EOB}
-\phi_{\rm metric}^{\rm NR}$. $\Delta \phi$ remains
remarkably small ($\sim \pm 0.02$ radians) during the
entire inspiral and plunge ($\omega_2=0.31$ being quite near
the merger, see inset). By comparison, the root-sum of the various
numerical errors on the phase 
(numerical truncation, outer boundary, 
extrapolation to infinity) is about $0.023$ radians
during the inspiral~\cite{Scheel:2008rj}. At the merger,
and during the ringdown, $\Delta \phi$ takes somewhat 
larger values ($\sim \pm 0.1$ radians), but it oscillates around
zero, so that, on average, it stays very well in phase
with the NR waveform (as is clear on Fig.~\ref{fig:fig1}). 
By comparison, we note that~\cite{Scheel:2008rj} mentions 
that the phase error linked to the extrapolation to infinity
doubles during ringdown. We also found that the NR
signal after merger is contaminated by unphysical oscillations.
We then note that the total ``two-sigma'' NR error level
estimated in~\cite{Scheel:2008rj} rises to
$ 0.05$ radians during ringdown, which is comparable 
to the EOB-NR phase disagreement. 
Figure~\ref{fig:fig3} compares the analytical and numerical 
metric moduli, $|\Psi_{22}|/\nu$. Again our (Pad\'e-resummed, 
NQC-corrected) analytical waveform yields a remarkably 
accurate description of the inspiral NR waveform. 
During the early inspiral the fractional agreement between 
the moduli is at the $3\times 10^{-3}$ level; as late as time 
$t=3900$, which corresponds  to 1.5 GW cycles before merger, 
the agreement is better than $1\times 10^{-3}$. 
The discrepancy between the two moduli starts being visible 
only just before and just after merger (where it remains at 
the $2.5\times 10^{-2}$ level). This very nice agreement 
should be compared with the previously considered EOB 
waveforms (which had a more primitive NQC factor, 
with $a_2=0$~\cite{Damour:2007vq,Damour:2008te})
which led to large moduli disagreements ($\sim 20 \%$,
see Fig.~9 in ~\cite{Damour:2008te}) at merger. 
By contrast, the present moduli disagreement is 
comparable to the estimated NR modulus fractional 
error (whose two-sigma level is $ 2.2\times10^{-2}$ 
after merger~\cite{Scheel:2008rj}).

\begin{figure}[t]
\begin{center}
\includegraphics[width=70 mm, height=55mm]{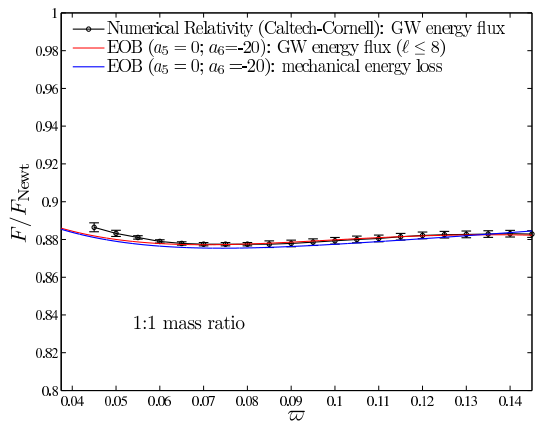}
\caption{ \label{fig:fig4} Triple comparison between NR and EOB GW energy fluxes and
  the EOB mechanical energy loss. }
\end{center}
\end{figure}

We also explored another aspect of the physical soundness
of our analytical formalism: the {\it triple} comparison
between (i) the NR GW energy flux at infinity
(which was computed in~\cite{Boyle:2008ge}); 
(ii) the corresponding analytically predicted 
GW energy flux at infinity (computed by summing 
$|{\dot h}_{\ell m}|^2$ over $\ell, m$ );
and (iii) (minus) the {\it mechanical} energy loss of the system,
as predicted by the general EOB formalism, i.e. the ``work''
done by the radiation reaction ${\dot E}_{\rm mechanical}
= \Omega {\mathcal F}_{\varphi}$. This comparison is
shown in Fig.~\ref{fig:fig4}, which should be compared 
to Fig. 9 of~\cite{Boyle:2008ge}. 
We kept the same vertical scale as~\cite{Boyle:2008ge} 
which compared the NR flux to older versions of 
(resummed and non-resummed) analytical fluxes 
and needed such a $\pm 10 \%$ vertical scale
to accomodate all the models they considered. 
[The horizontal axis is the frequency $\varpi$
of the differentiated metric waveform $\dot{h}_{22}$.] 
By contrast, we see again
the striking closeness (at the $ \sim 2 \times 10^{-3}$ level)
between the analytical and NR GW fluxes. As both fluxes
include higher multipoles than the $(2,2)$ one, this
closeness is a further test of the agreement between
our analytical formalism and NR results. [We think that
the $\sim 2 \sigma$ difference between the
(coinciding) analytical curves and the NR one on
the left of the Figure is due to uncertainties in the
flux computation of~\cite{Boyle:2008ge}, possibly related 
to their method of computing $\dot{h}$.]
Note that the rather close agreement between the analytical
energy flux and the mechanical energy loss during late
inspiral is not required by physics (because of the well-known
``Schott term''~\cite{Schott}), but is rather an 
indication that $\dot h_{\ell m}$ can be well
approximated by $- im \Omega h_{\ell m}$ 
(used in Eq.~\Ref{RR}).

\begin{figure}[t]
\begin{center}
\includegraphics[width=70 mm, height=55mm]{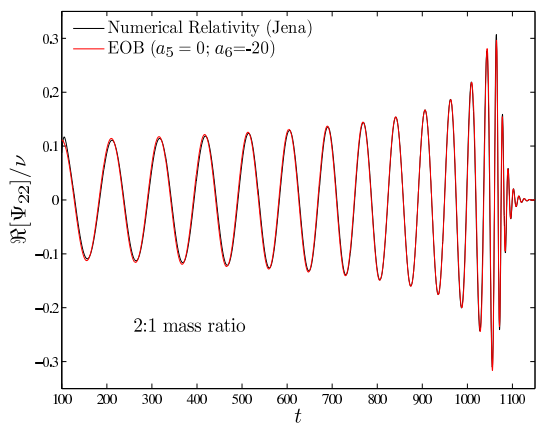}
\caption{ \label{fig:fig5}Unequal mass case:  Comparison between 
metric waveforms for the 2:1 mass ratio. }
\end{center}
\end{figure}
Finally, as the power of our formalism resides in the ease
with which it can accomodate continuous variations
in the basic physical parameters of the considered BBH,
we shall discuss an unequal-mass system ($\nu < 1/4$).
The highest-accuracy data that we had in hands is the
Jena-group simulation of a 2:1 mass ratio BBH ($\nu = 2/9=
0.22222$). When $a_5=0, a_6=-20$, and $\nu=2/9$, one finds
$a_1=-0.017017, a_2=1.1906$.
Using the data reported in~\cite{Damour:2008te} (and the
function $\varphi(\nu)$ quoted above), we
compare in Fig.~\ref{fig:fig5} 
(the real part of) our analytical {\it metric}
quadrupolar waveform $\Psi^{\rm EOB}_{22}/\nu$ to the
corresponding (Jena, 2:1 mass ratio) NR 
{\it metric} waveform $\Psi^{\rm NR}_{22}/\nu$.
[We use, as in~\cite{Damour:2008te},
the pinching frequencies $\omega_1=0.1005,
\omega_2=0.4542$ .]
Again we have an excellent analytical-numerical
agreement, both in phase and in modulus. The
small differences between the two are within the
numerical errors (see~\cite{Damour:2008te}).

{\it Conclusions.} We have described a specific analytical
formalism (within the EOB framework), which contains as
arbitrariness only the resummation-flexibility parameters
of the crucial EOB $A(u)$ potential. We have shown that
for certain values~\footnote{We leave a detailed discussion
of the ``degenerate set'' of ``good'' $(a_5,a_6)$  values to
a future publication~\cite{DN}.} of these parameters ($a_5,a_6$):
(i) the waveform predicted by our analytical formalism agrees,
essentially within numerical errors, with the currently  most
accurate numerical relativity simulations; this agreement
holds for several different mass ratios (1:1, 2:1 and 4:1 -- not shown here); and
(ii) the gravitational wave energy flux predicted by our
formalism agrees, within numerical errors, with the most
accurate numerical-relativity energy flux.
We think that our formalism  (possibly after some
further minor improvements) opens a realistic possibility
of constructing (with minimal computational resources) a
very accurate, large bank of gravitational wave templates,
thereby helping in both detecting and analyzing the
signals emitted by inspiralling and coalescing binary
black holes. [Though we have in mind essentially 
ground-based detectors, we think our method could also apply to
space-based ones.] Finally, from a theoretical point of view, we think that
our method can be extended to the description of
(nearly circularized) {\it spinning} black hole 
systems (see~\cite{Damour:2001tu}).

After the submission of this work, 
a paper~\cite{Buonanno:2009qa} 
comparing Caltech-Cornell
numerical data to a different version of the EOB 
formalism appeared on the archives.
The EOB formalism of~\cite{Buonanno:2009qa} does not
use our novel (predictive) radiation reaction 
Eq.~\eqref{RR} (but rather the $v_{\rm pole}$-tuned 
one advocated in \cite{Damour:2007yf}),
nor our $a_6$-improved $A$ potential, Eq.~\eqref{A}.
Moreover, by contrast to the approach advocated here to
reduce to an absolute minimum the number of adjusted 
parameters, namely two, ($a_5$, $a_6$), 
Ref.~\cite{Buonanno:2009qa} tunes six parameters:
($a_5(1/4)$, $v_{\rm pole}(1/4)$, $a_3^{h_{22}}(1/4)$, 
$a_4^{h_{22}}(1/4)$, $a_5^{h_{22}}(1/4)$, and $t^{22}_{\rm match}(1/4)$).

\end{document}